\documentclass[aps,preprint]{revtex4}%
\usepackage{amsfonts}
\usepackage{amsmath}
\usepackage{amssymb}
\usepackage{graphicx}
\usepackage{color}%
\providecommand{\U}[1]{\protect\rule{.1in}{.1in}}
\begin{document}
\title[Cosmic acceleration from second order gauge gravity]{Cosmic acceleration from second order gauge gravity}
\author{R. R. Cuzinatto$^{1}$\thanks{cuzinatto@gmail.com}}
\author{C. A. M. de Melo$^{1,2}$\thanks{cassius.anderson@gmail.com}}
\author{L. G. Medeiros$^{3}$\thanks{leogmedeiros@gmail.com}}
\author{P. J. Pompeia$^{4}$\thanks{ppompeia@phys.ualberta.ca. On leave from: Instituto
de F\'{\i}sica Te\'{o}rica, UNESP, S\~{a}o Paulo, Brazil.} }
\affiliation{$^{1}$Instituto de Ci\^{e}ncia e Tecnologia,}
\affiliation{Universidade Federal de Alfenas, Campus Po\c{c}os de Caldas.}
\affiliation{Rodovia Pref. Jos\'{e} Aur\'{e}lio Vilela (BR 267), Km 533, n$%
%TCIMACRO{\U{b0}}%
%BeginExpansion
{{}^\circ}%
%EndExpansion
$11999, CEP 37701-970, Po\c{c}os de Caldas, MG, Brazil.}
\affiliation{$^{2}$Instituto de F\'{\i}sica Te\'{o}rica, Universidade Estadual Paulista.}
\affiliation{Rua Bento Teobaldo Ferraz 271 Bloco II, P.O. Box 70532-2, CEP 01156-970,
S\~{a}o Paulo, SP, Brazil.}
\affiliation{$^{3}$Escola de Ci\^{e}ncia e Tecnologia, Universidade Federal do Rio Grande
do Norte. }
\affiliation{Campus Universit\'{a}rio, s/n - Lagoa Nova, Natal, RN, Brazil.\\}
\affiliation{$^{4}$Departamento de Ci\^{e}ncia e Tecnologia Aeroespacial, Instituto de
Fomento e Coordena\c{c}\~{a}o Industrial. }
\affiliation{Pra\c{c}a Mal. Eduardo Gomes 50, CEP 12228-901, S\~{a}o Jos\'{e} dos Campos,
SP, Brazil.}

\begin{abstract}
We construct a phenomenological theory of gravitation based on a second order
gauge formulation for the Lorentz group. The model presents a long-range
modification for the gravitational field leading to a cosmological model
provided with an accelerated expansion at recent times. We estimate the model
parameters using observational data and verify that our estimative for the age
of the Universe is of the same magnitude than the one predicted by the
standard model. The transition from the decelerated expansion regime to the
accelerated one occurs recently (at $\sim9.3\;Gyr$).

\end{abstract}
\maketitle

\section{Introduction}

One of the most challenging problems of Physics nowadays is to explain the
origin and evolution of the present accelerated expansion of the universe. One
way of obtaining a mechanism of acceleration is to modify one of the
cornerstones of modern physics, the theory of General Relativity.

Modifications in the scheme of General Relativity are being proposed since its
invention, in the beginning of the 20th century, and they are motivated by
several reasons, from the quest for agreement with the theory for the inner
structure of quantized matter, to the eventual need for extra-dimensions and
the desire to obtain unification of the interactions. The first modification
of General Relativity was proposed by Einstein through the introduction of the
cosmological constant, which is one of the several alternatives to describe
the present-day acceleration of the universe. Other proposals associated with
renormalizability are the quadratic Lagrangians in the Riemann tensor
\cite{RevR2} and the Horava-Lifshitz model \cite{Horava}; in the context of
Cosmology possible modifications involve the $f\left(  R\right)  $ Lagrangians
\cite{GRG, fR, Amendola}, the introduction of one (or more) spatial
extra-dimension in the braneworld scenario \cite{RevBraneWorld}, or the
presence of a self-interacting scalar field, the quintessence models
\cite{RevQuintessence}.

Another class of modified gravity theories consists of the inclusion of
non-local terms in the gravitational Lagrangians. Non-local terms arise
naturally, for instance, if one considers the inverse d 'Alembertian operator
\cite{Woodard}. This have inspired non-local modifications in $f\left(
R\right)  $ gravity which apply inverse differential operators to the Ricci
scalar \cite{PLB2008} or to the Gauss-Bonnet invariant \cite{PLB2009}.
Non-local theories can generate late time acceleration in the universe or even
flat rotation curves in galaxies \cite{Friedmann}.

Here, we shall explore some of the cosmological consequences of a
phenomenological theory of gravitation based on a Lagrangian analyzed in
\cite{EPJC}; we will show that this theory permits a recent accelerated phase
for the universe without the introduction of $\Lambda$, exotic matter,
extra-dimensions or scalar fields. It was constructed on the basis of a gauge
formulation for the gravitational field, through the second order gauge theory
\cite{2ndOrd}. In the gauge approach to gravity the simplest choice for the
local gauge symmetry is the Lorentz group. If one considers the second order
extension of the Lorentz gauge theory, the gravitational field Lagrangian must
depend on the second derivative of the spin connection. Therefore, in the
geometrical framework, the second order extension of the Einstein gravity
should consider the gradient of the scalar curvature rather then a quadratic
term in the curvature. The second order gauge theory was shown to be efficient
in the description of effective limits of other gauge theories, such as
Podolsky electrodynamics and the $SU(N)$ non-abelian model.

From the gauge theoretical point of view, the Einstein-Hilbert action, the
quadratic Lagrangians in the Riemann tensor and the $f\left(  R\right)  $
Lagrangians are all of first order, as discussed in \cite{EPJC}, where it was
performed a classification of all possible quadratic Lagrangians of first and
second order in the gauge gravitational field. Among these possibilities, we
chose a particularly simple one, which is inspired by the Podolsky's abelian
case and by the effective Alekseev-Arbuzov-Baikov's non-abelian model
\cite{AAB}. In fact, we have added a term scaling with the square of the
covariant derivative of the scalar curvature $\nabla R$ to the familiar $R$ of
the Einstein-Hilbert Lagrangian. Our main intention was to analyze the
consequences of this option in the cosmological context and how the higher
order derivative terms could reproduce the present-day cosmic acceleration of
the scale factor.

We will adopt the following action for the description of the gravitational
field:
\begin{align}
&  \left.  S=\int d^{4}x\sqrt{-g}\left(  \frac{R}{2\chi}+\frac{\beta}{\chi
}\mathcal{L}_{P}-\mathcal{L}_{M}\right)  ~,\right. \label{AcEff}\\
&  \left.  \mathcal{L}_{P}=\frac{1}{8}\nabla_{\mu}R\nabla^{\mu}R~,\right.
\label{PodolskyGrav}%
\end{align}
and derive the field equations using the classical Schwinger-Weiss variational
principle. The Schwinger action principle was introduced in the context of
Quantum Field Theory \cite{QuanField} and recently has been used to study
classical and quantum fields in spacetimes  with curvature and torsion
\cite{CQG} or even to investigate the gauge fixing in quantized
electromagnetic field \cite{Bfield}.

It is worth to emphasize that the choice (\ref{PodolskyGrav}) constitutes a
phenomenological model valid within a limited interval of energy (set by the
values of the coupling constant $\beta$); it does not hold during all the
cosmological history (as we shall see) but only for a certain period. The same
phenomenological Lagrangian was used by Gottl\"{o}ber \emph{et al.} in another
context \cite{Starobinski}, where the authors claimed that the consideration
of this higher order term \textquotedblleft could be thought of as an attempt
to make a further step in understanding the features of (...) non-local
interaction\textquotedblright.

The paper is organized as follows. In section \ref{FriedEq} the field
equations are written for a Friedmann-Lema\^{\i}tre-Robertson-Walker metric.
Section \ref{Solution} is devoted to obtain a perturbative solution of the
field equations about the usual dust-matter model of the Einstein-Hilbert
theory (as described by the Friedmann equations). The perturbative solution is
constructed in such a way that the universe is dominated by a decelerated
regime until the time $t^{\ast}$ when the additional term $\frac{\beta}{\chi
}\mathcal{L}_{P}$ begins to be relevant. In section \ref{Observations} the
parameters of the model are related to the observational data through a set of
coupled nonlinear equations. Such equations are solved by numerical methods in
section \ref{Numbers}, and the results are discussed in section
\ref{Conclusion}.

\section{Friedmann Equations\label{FriedEq}}

The invariance of the action (\ref{AcEff}) with respect to $\delta
g_{\lambda\nu}$ yields the field equations:
\begin{align}
&  \left.  R_{\lambda\nu}-\frac{1}{2}g_{\lambda\nu}R+\beta^{2}H_{\lambda\nu
}=\chi T_{\lambda\nu}~,\right. \label{FieldEq}\\
&  \left.  H_{\lambda\nu}=\nabla_{\lambda}\nabla_{\nu}\left[  \square
R\right]  +\frac{1}{2}\nabla_{\lambda}R\nabla_{\nu}R-R_{\lambda\nu}\square
R-g_{\lambda\nu}\square\left[  \square R\right]  -\frac{1}{4}g_{\lambda\nu
}\nabla^{\rho}R\nabla_{\rho}R\right.  ~.\nonumber
\end{align}
where $\square\equiv\nabla_{\mu}\nabla^{\mu}$ and $\nabla_{\mu}$ is the
covariant derivative.

Applying the field equations (\ref{FieldEq}) to a homogeneous and isotropic
space, described by the Friedmann-Lema\^{\i}tre-Robertson-Walker (FLRW)
metric,%
\[
ds^{2}=dt^{2}-a^{2}\left(  t\right)  \left(  \frac{1}{1-\kappa r^{2}}%
dr^{2}+r^{2}d\Omega^{2}\right)
\]
$\left(  \kappa=-1,0,+1\right)  $\ one finds, after some direct but long
calculations,%
\begin{align}
&  \left.  -3\left(  \dot{H}+H^{2}\right)  -\frac{1}{2}R+\beta\left[
-3H\dddot{R}+3\dot{H}\ddot{R}-6H^{2}\ddot{R}+9H^{3}\dot{R}+\frac{1}{4}\dot
{R}^{2}\right]  =\chi T_{00}\right.  ~,\nonumber\\
&  \left.  \frac{a^{2}}{\left(  1-\kappa r^{2}\right)  }\left[  \dot{H}%
+3H^{2}+\frac{1}{2}R+2\frac{\kappa}{a^{2}}+\beta\left(  \ddddot{R}+5H\dddot
{R}+3H^{2}\ddot{R}+5\dot{H}\ddot{R}+\right.  \right.  \right. \nonumber\\
&  \left.  \left.  \left.  +3H\dot{H}\dot{R}-9H^{3}\dot{R}+3\ddot{H}\dot
{R}+\frac{1}{4}\dot{R}^{2}-2\frac{\kappa}{a^{2}}\left(  \ddot{R}+3H\dot
{R}\right)  \right)  \right]  =\chi T_{11}\right.  ~. \label{FriedFieldEq}%
\end{align}
where $H\left(  t\right)  =\dot{a}/a$ is the Hubble function, $R\left(
t\right)  =g^{\mu\nu}R_{\mu\nu}$ is the scalar curvature and we are using
units such that $\chi=8\pi G$. In our notation, dot\ means derivation with
respect to the cosmic time $t$. These are the higher order Friedmann equations
in terms of the Hubble function $H\left(  t\right)  $ and the scalar curvature
$R\left(  t\right)  $.

Following the standard procedure we use the energy-momentum tensor of a
perfect fluid in a commoving coordinate system,
\[
T_{\mu\nu}=\left(  \rho+p\right)  \delta_{\mu}^{0}\delta_{\nu}^{0}-pg_{\mu\nu
}~.
\]

In order to simplify the treatment, we will be concerned only with the case of
a flat spatial section, $\kappa=0$. So, using the relationship between the
scalar curvature and the Hubble function,
\[
R=-6\left(  \dot{H}+2H^{2}\right)  ~,
\]
we get the following modified Friedmann equations:
\begin{align*}
&  3H^{2}+\beta\left(  18H\ddddot{H}+108H^{2}\dddot{H}-18\dot{H}\dddot
{H}+9\ddot{H}^{2}+90H^{3}\ddot{H}+\right. \\
&  \left.  \left.  +216H\dot{H}\ddot{H}-72\dot{H}^{3}+288\left(  H\dot
{H}\right)  ^{2}-216H^{4}\dot{H}\right)  =\chi\rho\right.  ~,\\
&  2\dot{H}+3H^{2}+\beta\left(  6H^{\left(  5\right)  }+54H\ddddot{H}%
+138H^{2}\dddot{H}+126\dot{H}\dddot{H}+\right. \\
&  \left.  \left.  +81\ddot{H}^{2}+18H^{3}\ddot{H}+498H\dot{H}\ddot{H}%
+120\dot{H}^{3}-216H^{4}\dot{H}\right)  =-\chi p\right.  ~.
\end{align*}

Combining the equations, one finds:
\begin{align}
&  2\dot{H}+\beta\left(  6H^{\left(  5\right)  }+36H\ddddot{H}+30H^{2}%
\dddot{H}+144\dot{H}\dddot{H}+72\ddot{H}^{2}+\right. \label{EqMotion}\\
&  \left.  \left.  -72H^{3}\ddot{H}+282H\dot{H}\ddot{H}+192\dot{H}%
^{3}-288\left(  H\dot{H}\right)  ^{2}\right)  =-\chi\left(  p+\rho\right)
\right.  ~.\nonumber
\end{align}

Once we want to describe the evolution of the universe, this equation must be
complemented with the covariant conservation of energy-momentum,
\[
\dot{\rho}+3H\left(  \rho+p\right)  =0~,
\]
and an equation of state $f$ relating the energy density $\rho$, the pressure
$p$ and the Hubble function $H$,
\[
f\left(  \rho,p,H\right)  =0~.
\]
The dependence on $H$ is included to account for the general case when one
admits interaction among the constituents of the cosmic fluid \cite{Leo}. In
this case, there is a possible constraint relating $p$, $\rho$ and the scale
factor, or equivalently $H$. On the other hand, the usual equations of state
of physical cosmology associate only pressure $p$ to the energy density $\rho
$, or pressure to the numerical density $n$. For example, the equation of
state for the dust matter is $p=nkT\ll\rho$ ($k$ is the Boltzmann constant and
$T$ the temperature), and $p=\rho/3$ is the one used for ultra-relativistic particles.

\section{Solutions of the higher order Friedmann equations\label{Solution}}

\subsection{Dust Matter}

Our main interest here is to apply the model to the present state of the
universe. Therefore, we take as source a perfect fluid composed by dust matter
$p=0$ (ordinary or dark). In this case, the continuity equation gives:
\[
\rho\left(  t\right)  =\rho_{0}\left(  \frac{a_{0}}{a\left(  t\right)
}\right)  ^{3}~.
\]

In order to use such result directly, we would have to rewrite equation
(\ref{EqMotion}) in terms of the scale factor, obtaining a nonlinear and much
more complicate equation, which we shall avoid. Instead, we consider
simultaneously the following pair of coupled equations:
\begin{align}
&  \left.  \dot{H}+\beta\left(  3H^{\left(  5\right)  }+18H\ddddot{H}%
+15H^{2}\dddot{H}+72\dot{H}\dddot{H}+36\ddot{H}^{2}+\right.  \right.
\nonumber\\
&  \left.  \left.  -36H^{3}\ddot{H}+141H\dot{H}\ddot{H}+96\dot{H}%
^{3}-144\left(  H\dot{H}\right)  ^{2}\right)  =-\frac{\chi}{2}\rho\right.
~,\label{SecOrdEq}\\
&  \left.  \dot{\rho}+3H\rho=0\right.  ~.\nonumber
\end{align}

These equations can be analyzed by several methods, such as the linearization
of dynamical systems, spectral analysis or perturbation theory. Here we will
consider only this last procedure, leaving the other options for future investigations.

\subsection{Perturbation Theory}

The model is constructed by assuming a standard Friedmann expansion prior to
some time $t^{\ast}$ from which the second order effects start to become
significant. The strategy is to consider a perturbation series in the coupling
parameter $\beta$, in order to guarantee the accordance of our model with the
usual cosmological model (in some region of the space of parameters). Take,
for instance, an expansion up to second order terms; it reads:%
\begin{align}
H\left(  t\right)   &  = H_{F}+\beta H_{1}+\beta^{2}H_{2}~,\nonumber\\
\rho\left(  t\right)   &  = \rho_{F}+\beta\rho_{1}+\beta^{2}\rho_{2}~,
\label{expansions}%
\end{align}
where the label $F$ stands for the standard Friedmann solution of the Einstein equations.

Substituting expansions (\ref{expansions}) in\ the pair (\ref{SecOrdEq}) and
matching the terms order by order, we get:%
\begin{equation}
\mathcal{O}\left(  \beta^{0}\right)  \rightarrow\left\{
\begin{array}
[c]{l}%
\dot{H}_{F}+\frac{\chi}{2}\rho_{F}=0\\
\dot{\rho}_{F}+3H_{F}\rho_{F}=0
\end{array}
\right.  ~, \label{O(beta0)}%
\end{equation}
and,%
\begin{equation}
\mathcal{O}\left(  \beta^{1}\right)  \rightarrow\left\{
\begin{array}
[c]{l}%
\dot{H}_{1}+\frac{\chi}{2}\rho_{1}=S_{1}\left(  t\right) \\
\dot{\rho}_{1}+3H_{F}\rho_{1}+3H_{1}\rho_{F}=0
\end{array}
\right.  ~, \label{O(beta1)}%
\end{equation}
where%
\begin{align}
S_{1}\left(  t\right)   &  \equiv-\left(  3H_{F}^{\left(  5\right)  }%
+18H_{F}\ddddot{H}_{F}+15H_{F}^{2}\dddot{H}_{F}+72\dot{H}_{F}\dddot{H}%
_{F}+36\ddot{H}_{F}^{2}\right. \label{S1}\\
&  \left.  -36H_{F}^{3}\ddot{H}_{F}+141H_{F}\dot{H}_{F}\ddot{H}_{F}+96\dot
{H}_{F}^{3}-144H_{F}^{2}\dot{H}_{F}^{2}\right)  ~;\nonumber
\end{align}
and also,%
\begin{equation}
\mathcal{O}\left(  \beta^{2}\right)  \rightarrow\left\{
\begin{array}
[c]{l}%
\dot{H}_{2}+\frac{\chi}{2}\rho_{2}=S_{2}\left(  t\right) \\
\dot{\rho}_{2}+3H_{F}\rho_{2}+3H_{2}\rho_{F}=-3H_{1}\rho_{1}%
\end{array}
\right.  ~, \label{O(beta2)}%
\end{equation}
with,%
\begin{align}
S_{2}\left(  t\right)   &  \equiv-\left[  3H_{1}^{\left(  5\right)
}+18\left(  H_{1}\ddddot{H}_{F}+H_{F}\ddddot{H}_{1}\right)  +30H_{F}%
H_{1}\dddot{H}_{F}+\right. \nonumber\\
&  \left.  +72\left(  \dot{H}_{F}\dddot{H}_{1}+\dot{H}_{1}\dddot{H}_{F}%
+\ddot{H}_{F}\ddot{H}_{1}\right)  -108H_{1}H_{F}^{2}\ddot{H}_{F}-36H_{F}%
^{3}\ddot{H}_{1}\right]  +\label{S2}\\
&  -\left[  141\left(  H_{1}\dot{H}_{F}\ddot{H}_{F}+H_{F}\dot{H}_{F}\ddot
{H}_{1}+H_{F}\dot{H}_{1}\ddot{H}_{F}\right)  +288\left(  \dot{H}_{1}\dot
{H}_{F}^{2}-H_{F}^{2}\dot{H}_{F}\dot{H}_{1}-H_{F}H_{1}\dot{H}_{F}^{2}\right)
\right]  ~.\nonumber
\end{align}

This way, we obtained a pair of coupled linear equations in each order. Their
previous orders give the source term and the coefficients.

\subsubsection{Zeroth order solution: the standard cosmological model}

The solution for the system of zeroth order in the coupling parameter $\beta$,
Eq. (\ref{O(beta0)}),%
\begin{align}
&  \left.  \dot{H}_{F}+\frac{\chi}{2}\rho_{F}=0~,\right. \label{Friedmann1}\\
&  \left.  \dot{\rho}_{F}+3H_{F}\rho_{F}=0~,\right.  \label{Friedmann2}%
\end{align}
can be obtained by direct integration. Solving this coupled system, we have:%
\begin{align}
H_{F}  &  =\frac{2}{3}\frac{1}{t}~,\label{HF}\\
\rho_{F}  &  =\frac{3}{\chi}H_{F}^{2}~,\nonumber
\end{align}
with an appropriate initial condition.

\subsubsection{First order solution}

In the first order approximation, we have the coupled set (\ref{O(beta1)}),%
\begin{align*}
&  \left.  \dot{H}_{1}+\frac{\chi}{2}\rho_{1}=S_{1}\left(  t\right)  ~,\right.
\\
&  \left.  \dot{\rho}_{1}+3H_{F}\rho_{1}+3H_{1}\rho_{F}=0~,\right.
\end{align*}
These equations can be solved by the Increasing Order Method. Differentiating
the first of these equations and using the second one, we obtain
\begin{align*}
&  \ddot{H}_{1}+3H_{F}\dot{H}_{1}-\frac{3\chi}{2}\rho_{F}H_{1}=\tilde{S}%
_{1}\left(  t\right) \\
&  \tilde{S}_{1}\left(  t\right)  =\dot{S}_{1}\left(  t\right)  +3H_{F}%
S_{1}\left(  t\right)
\end{align*}
The general solution of such equation can be obtained in the form of a power
law:%
\begin{align*}
H_{1}\left(  t\right)   &  =at+bt^{-2}+\frac{4912}{243}t^{-5}\\
\rho_{1}\left(  t\right)   &  =\frac{2}{\chi}\left(  -\frac{15\,977}%
{243}t^{-6}-a+2bt^{-3}\right)
\end{align*}

The integration constants $a$ and $b$\ should be chosen in accordance to the
physical situation to be described. Since the zeroth order terms appear as
source terms in the first order approximation, one can choose the initial
conditions $H_{1}\left(  t^{\ast}\right)  = \rho_{1} \left(  t^{\ast}\right)
= 0$. This determines the integration constants leaving the theory with only
three free parameters, namely the coupling constant $\beta$, the age of the
universe $t_{0}$ (see below) and the instant of perturbation $t^{\ast}$.

Therefore, in the first order approximation we find the following solution to
equations (\ref{SecOrdEq}):%
\begin{align}
&  \left.  H\left(  t\right)  =\frac{2}{3}\frac{1}{t}+\frac{\beta}{\left(
t^{\ast}\right)  ^{4}}\left(  \frac{11\,065}{729}\left(  \frac{t^{\ast}}%
{t}\right)  t^{-1}+\frac{4912}{243}\left(  \frac{t^{\ast}}{t}\right)
^{4}t^{-1}-\frac{35\,408}{729}\frac{t}{t^{\ast}}\left(  t^{\ast}\right)
^{-1}\right)  \right. \nonumber\\
&  \left.  8\pi G\rho\left(  t\right)  =\frac{4}{3}\frac{1}{t^{2}}%
+2\frac{\beta}{\left(  t^{\ast}\right)  ^{4}}\left(  \frac{22\,130}%
{729}\left(  \frac{t^{\ast}}{t}\right)  t^{-2}-\frac{15\,977}{243}\left(
\frac{t^{\ast}}{t}\right)  ^{4}t^{-2}+\frac{35\,408}{729}\left(  t^{\ast
}\right)  ^{-2}\right)  \right.  ~. \label{system H rho}%
\end{align}

\section{Observational Parameters\label{Observations}}

Now, let us focus on the problem of estimating the magnitude of the parameters
of our theoretical model using the observational data available.

The redshift $z$ and the luminosity distance $d_{L}$ are dependent on the null
geodesic equation only, and this is not changed by the second order field
equations (\ref{FieldEq}). Therefore, they constitute the ideal data set to be
compared with the predictions of our model. The luminosity distance\ can be
directly related to the redshift \cite{Weinberg},
\[
d_{L}\approx\frac{1}{H_{0}}\left(  z+\frac{1}{2}\left(  1-q_{0}\right)
z^{2}\right)  ~.
\]
$q_{0}$ is the deceleration parameter.

The supernovae projects usually measure the curve of $d_{L}\left(  z\right)  $
determining the parameter $q_{0}$ with good accuracy. In order to do it,
supernovae projects maximize the likehood function adjusting the model
parameters. Instead of following this approach, we shall obtain an initial
estimation for our free parameters by searching for values which are as
independent of model as possible.

In our model we have three parameters to be found:

\begin{enumerate}
\item The age of the universe $t_{0}$;

\item The instant $t^{\ast}$ from which the perturbation coming from the
modified gravitational equation becomes important;

\item The coupling constant $\beta$ for the higher derivative terms in the action.
\end{enumerate}

We need three independent measurements to find these parameters. We will use
$H_{0}$, $q_{0}$ and $\Omega_{m0}$ obtained from the literature.

In the following section, we shall carefully discuss how to use $H_{0}$,
$q_{0}$\ and $\Omega_{m}$ to obtain $t_{0}$, $t^{\ast}$\ and $\beta$. But,
before that, we will add to system (\ref{system H rho}) the constraint
\begin{equation}
\dot{H}\left(  t_{0}\right)  =-H_{0}^{2}\left(  q_{0}+1\right)  \label{H dot}%
\end{equation}
following from the definition of both the Hubble and the deceleration
functions in terms of the scale factor: $H=\dot{a}/a$, $q=-\ddot{a}a/\dot
{a}^{2}$. Gathering (\ref{system H rho}) and (\ref{H dot}), we get the new
system to be solved:
\begin{align*}
H\left(  t_{0}\right)   &  =\frac{2}{3}\frac{1}{t_{0}}+\frac{\beta}{\left(
t^{\ast}\right)  ^{4}}\left(  -\frac{35\,408}{729}\frac{t_{0}}{t^{\ast}%
}\left(  t^{\ast}\right)  ^{-1}+\frac{11\,065}{729}\left(  \frac{t_{0}%
}{t^{\ast}}\right)  ^{-1}t_{0}^{-1}+\frac{4912}{243}\left(  \frac{t_{0}%
}{t^{\ast}}\right)  ^{-4}t_{0}^{-1}\right) \\
\dot{H}\left(  t_{0}\right)   &  =-\frac{2}{3}\frac{1}{t_{0}^{2}}+\frac{\beta
}{\left(  t^{\ast}\right)  ^{4}}\left(  -\frac{35\,408}{729}\left(  t^{\ast
}\right)  ^{-2}-\frac{22\,130}{729}\left(  \frac{t_{0}}{t^{\ast}}\right)
^{-1}t_{0}^{-2}-\frac{24\,560}{243}\left(  \frac{t_{0}}{t^{\ast}}\right)
^{-4}t_{0}^{-2}\right) \\
8\pi G\rho\left(  t_{0}\right)   &  =\frac{4}{3}\frac{1}{t_{0}^{2}}%
+2\frac{\beta}{\left(  t^{\ast}\right)  ^{4}}\left(  -\frac{15\,977}%
{243}\left(  \frac{t_{0}}{t^{\ast}}\right)  ^{-4}t_{0}^{-2}+\frac
{35\,408}{729}\left(  t^{\ast}\right)  ^{-2}+\frac{22130}{729}\left(
\frac{t_{0}}{t^{\ast}}\right)  ^{-1}t_{0}^{-2}\right)
\end{align*}
The first member of each equation of the system above is given in terms of
observational constants, while the right hand side of each equality bears the
parameters of the perturbed model. These will be calculated by solving
numerically the above transcendental equations. We deal with this task now.

\section{Numerical Calculations\label{Numbers}}

In order to solve numerically the system of coupled transcendental equations,
let us perform some simple manipulations. First, we define new non-dimensional
variables,
\begin{equation}
u\equiv\frac{t^{\ast}}{t_{0}}~,\;b\equiv\frac{\beta}{t^{\ast4}}~; \label{u b}%
\end{equation}
in terms of which the system is rewritten as%
\begin{align}
H_{0}  &  = \frac{2}{3}\frac{1}{t_{0}}\left(  1+b\left(  -\frac{17\,704}
{243}\frac{1}{u^{2}}+\frac{11\,065}{486}u+\frac{2456}{81}u^{4}\right)
\right)  \, ;\label{Hzero}\\
H_{0}^{2}\left(  q_{0}+1\right)   &  = \frac{2}{3}\frac{1}{t_{0}^{2}}\left(
1+b\left(  \frac{17\,704}{243}\frac{1}{u^{2}}+\frac{11\,065}{243}
u+\frac{12\,280}{81}u^{4}\right)  \right)  \, ;\label{HdotZero}\\
3H_{0}^{2}\Omega_{m0}  &  = \frac{4}{3}\frac{1}{t_{0}^{2}}\left(  1+b\left(
-\frac{15\,977}{162}u^{4}+\frac{17\,704}{243}\frac{1}{u^{2}}+\frac
{11\,065}{243}u\right)  \right)  \, , \label{OmegaM}%
\end{align}
where
\[
\Omega_{m0}\equiv\frac{8\pi G}{3H_{0}^{2}}\rho_{0} \, .
\]

Taking the ratio of the two last equations of the system above, one gets:
\begin{gather}
b\left(  u\right)  =\frac{2\left(  q_{0}+1\right)  -3\Omega_{m0}}{s\left(
u\right)  } \, ,\label{B}\\
s\left(  u\right)  \equiv\left(  3\Omega_{m0}-2\left(  q_{0}+1\right)
\right)  \left(  \frac{17\,704}{243}\frac{1}{u^{2}}+\frac{11\,065}{243}
u+\frac{36\,840}{243}u^{4}\right)  +\frac{40\,537}{81}\left(  q_{0}+1\right)
u^{4} \, .\nonumber
\end{gather}
We must have $\beta<0$ in order to assure the stability of the theory
\cite{EPJC, Starobinski}. This establishes a constraint on the sign of $b$.
Substituting $b\left(  u\right)  $ in the equation for $H_{0}$:
\begin{equation}
\frac{1}{t_{0}\left(  u\right)  }=\frac{3}{2}\frac{H_{0}}{\left(  1+b\left(
u\right)  \left(  -\frac{17\,704}{243}\frac{1}{u^{2}}+\frac{11\,065}%
{486}u+\frac{2456}{81}u^{4}\right)  \right)  } \, . \label{tzero}%
\end{equation}

Combining (\ref{Hzero}) and (\ref{HdotZero}) we get
\[
H_{0}^{2}\left(  2\left(  q_{0}+1\right)  -3\Omega_{m0}\right)  =\frac{4}%
{3}\frac{b\left(  u\right)  }{t_{0}^{2}\left(  u\right)  }\left(
\frac{12\,280}{81}+\frac{15\,977}{162}\right)  u^{4}\,,
\]
which is a nonlinear equation for the parameter $u$. This equation can be
solved using the Newton-Raphson method. The result is then used in (\ref{B})
to obtain the numerical value of $b$ and both $b$ and $u$ are inserted in
(\ref{tzero}) to get the age of the universe $t_{0}$. The time of perturbation
$t^{\ast}$ can then be obtained as a simple ratio. Proceeding this way, we are
able to find all the parameters of the model given the measurements of $H_{0}%
$, $q_{0}$ and $\Omega_{m0}$ (or $\rho_{0}$).

The Hubble constant is measured with great accuracy by the Hubble Space
Telescope Key Project \cite{HubbleKey}. The deceleration parameter is taken
from Rapetti \emph{et al.} \cite{Jerk} who used a kinematical approach to
cosmological expansion. The matter density $\Omega_{m0}$ was evaluated
following the same procedure as \cite{Allen} but retaining just the three
galaxy clusters with redshift below $0.1$ since this minimizes the dependence
of the estimation of $\Omega_{m0}$ on different cosmological models.

With this assumptions and using the experimental values
\[
H_{0} =0.074~\left(  Gyr\right)  ^{-1}, \, \, \, \, \, q_{0} =-0.81, \, \, \,
\, \, \Omega_{m0} =0.276~,
\]
we find:
\[
u \approx0.75, \, \, \, \, \, b \approx-0.0036, \, \, \, \, \, t^{\ast}
\approx9.3\;Gyr~,
\]
which can be used to estimate the parameter $\beta$ and the age of the
universe,
\begin{align}
\beta &  \approx-26\;\left(  Gyr\right)  ^{4}\,,\label{beta}\\
t_{0}  &  \approx12.4\;Gyr~. \label{t0}%
\end{align}

By varying $q_{0}$ in $20\%$ we obtain numerically a variation less than $2\%$
for $t_{0}$ and less than $20\%$ for $t^{\ast}$ what shows that the model is
aproximatelly robust.

\begin{figure}[h]
\begin{center}
\includegraphics[width=10cm]{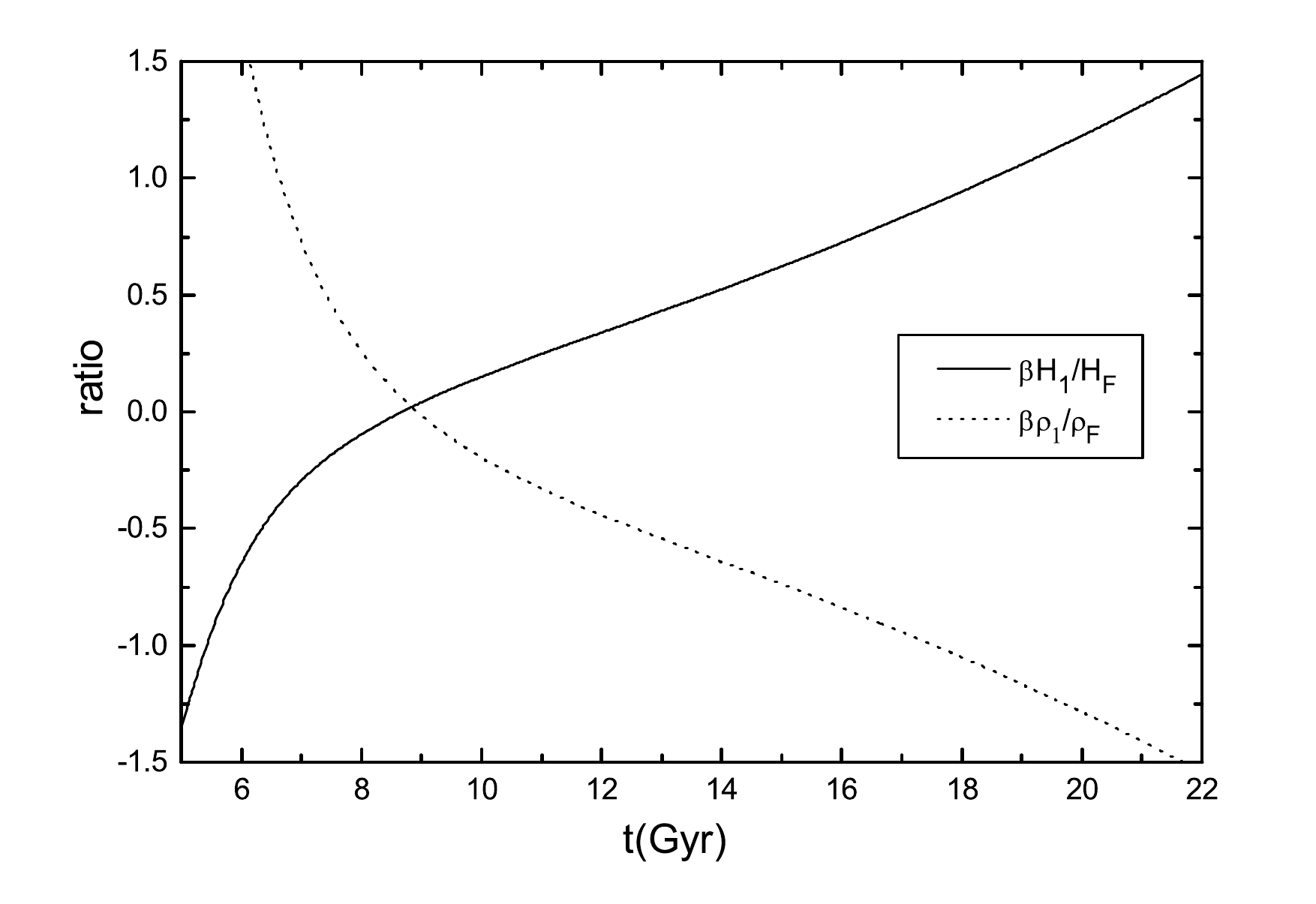}
\end{center}
\caption{Ratios of the first term in the perturbative solutions by the
ordinary dust-matter solution as functions of the time. The full line shows
the evolution of the the ratio of the first-order perturbed Hubble function
divided by the non-perturbed one; the dashed line presents the behavior of the
perturbation on the energy density divided by its non-perturbed values.}%
\label{fig2}%
\end{figure}

In spite of its apparently high value, $\beta$ does not break the meaning of
the modified action as proposed in (\ref{AcEff}). This is so because the first
order perturbation is such that the term $\beta H_{1}$ scales as
$\beta/t^{\ast4}$ and therefore is small when compared to the usual Friedmann
term $H_{F}$. Of course, the contribution of the additional term grows for
values of $t$ progressively greater than $t^{\ast}$. But we do not hope that
the perturbative approach holds for arbitrary large values of $t$. The
convergence of the perturbative expansion can be qualitatively studied
plotting the ratio of the perturbation term by the usual FLRW solution, as
given in Fig. \ref{fig2}.

%%%%%%%%%%%%%%%%%%%%%%%%%%%%
%%%%%%%%%% FIGURA %%%%%%%%%%
%%%%%%%%%%%%%%%%%%%%%%%%%%%%

%%%%%%%%%%%%%%%%%%%%%%%%%%%%

As long as the curves lie within the interval $\left[  -1,1\right]  $ the
perturbative scheme can be assumed as valid. Notice that the Hubble function
fits the perturbative scheme for a longer time compared to the behavior of the
energy density.

The value of $t_{0}$ provided by the WMAP3 data \cite{PDG} -- which assumes
the $\Lambda CDM$\ model as the supernovae approach does -- is calculated as
$13.7_{-0.2}^{+0.1}\;Gyr$ (for a flat Universe). The result given by our model
is roughly close with the one predicted by the $\Lambda CDM$ model. This
apparent tension between both results does not impair the model proposed here
since we are just looking for a preliminar estimation of our free parameters.
Besides, globular clusters data gives $11.2$ $Gyr$, at $95\%$ of confidence
level, as inferior limit for the age of the universe \cite{Krauss}.

Using the solution (\ref{system H rho}), one easily finds the ratio of scale
factor at two arbitrary times $t_{i}$ and $t_{f}$ as
\begin{gather*}
\ln\frac{a_{f}}{a_{i}}=\ln\left(  \frac{t_{f}}{t_{i}}\right)  ^{\frac{2}{3}%
}-\frac{\beta}{\left(  t^{\ast}\right)  ^{4}}\left(  \frac{17\,704}%
{729}\left(  \frac{t_{f}}{t^{\ast}}\right)  ^{2}+\frac{11\,065}{729}\left(
\frac{t_{f}}{t^{\ast}}\right)  ^{-1}+\frac{1228}{243}\left(  \frac{t_{f}%
}{t^{\ast}}\right)  ^{-4}\right)  +\\
+\frac{\beta}{\left(  t^{\ast}\right)  ^{4}}\left(  \frac{17\,704}{729}\left(
\frac{t_{i}}{t^{\ast}}\right)  ^{2}+\frac{11\,065}{729}\left(  \frac{t_{i}%
}{t^{\ast}}\right)  ^{-1}+\frac{1228}{243}\left(  \frac{t_{i}}{t^{\ast}%
}\right)  ^{-4}\right)  ~,
\end{gather*}
which exhibits a very smooth transition from the Friedmann standard regime to
the accelerated one.

The same quantity can be used to estimate the red-shift at the transition,%
\[
1+z^{\ast}=\frac{a_{0}}{a\left(  t^{\ast}\right)  }~,
\]
as%
\[
1+z^{\ast}=\left(  \frac{t_{0}}{t^{\ast}}\right)  ^{\frac{2}{3}}\times
\exp\left(  -\frac{\beta}{\left(  t^{\ast}\right)  ^{4}}\left(  \frac
{17\,704}{729}\left(  \frac{t_{0}}{t^{\ast}}\right)  ^{2}+\frac{11\,065}%
{729}\left(  \frac{t_{0}}{t^{\ast}}\right)  ^{-1}+\frac{1228}{243}\left(
\frac{t_{0}}{t^{\ast}}\right)  ^{-4}-\frac{32\,453}{729}\right)  \right)  ~.
\]
With the values of $\beta$, $t^{\ast}$ and $t_{0}$, one calculates $z^{\ast
}\approx0.27~$.

\section{Conclusions\label{Conclusion}}

We have constructed a model based on a phenomenological theory of gravitation
obtained from the inclusion of a Podolsky-like term scaling with the square of
the covariant derivative of the Ricci scalar. This model implies long-range
modifications in gravitation, which leads to an accelerated regime for the
present-day universe, even in the absence of a dark energy component or
cosmological constant. According to our perturbative evaluation, this
accelerated expansion started recently as indicated by the values of $t^{\ast
}$ or, equivalently, $z^{\ast}$.

The estimations given by the model for the age of the universe and the
redshift of transition are close to the supernovae data \cite{Supernova} or
the analysis of the cosmic microwave background based on the $\Lambda CDM$
model \cite{WMAP}. Other modified gravity theories -- e.g., the $f\left(
R\right)  $ theories \cite{GRG} -- can generate accelerated phases for the
expansion of the universe. In this paper, we obtained the same qualitative
results by adding a term proportional to $\left(  \nabla R\right)  ^{2}$ to
the Einstein-Hilbert Lagrangian. Our future perspectives include to obtain
more accurate values for the parameters using the likelihood function fitted
by supernovae data. We also intend to study a perturbative solution for a
closed $\left(  \kappa=1\right)  $ universe, keeping in mind the exact
solution found in the ordinary FLRW case \cite{Solucoes}.

The same phenomenological Lagrangian presented here was applied to describe
inflation in Ref. \cite{Starobinski}, but we emphasize the high order of
magnitude of the energies involved there, which would correspond to the early
stages of evolution of the Universe. On the other hand, the model presented
here engenders sensible dynamical effects at recent cosmic times, where the
energy scale is very low.

The fact that the gravitational field is weaker at long distances with a
characteristic scale given by the coupling constant $\beta$ suggests the
existence of massive modes in the weak field approximation, but in a way that
does not break the coordinate invariance, analogously to what happens in the
Podolsky electrodynamics \cite{2ndOrd}. The eventual existence of such massive
modes are under investigation, and the results should be compared to other
approaches in the same direction \cite{MassGravNovello}.

\acknowledgments

This work was partially supported by FAPEMIG-Brazil. The authors thank
Instituto de F\'{\i}sica Te\'{o}rica, Universidade Estadual Paulista, Brazil,
for providing the facilities. RRC, CAMM and LGM are grateful to the ICRA-BR
and the organizing committee of the Brazilian School of Cosmology and
Gravitation for the stimulating environment where this work was initiated.

\end{document}